\documentstyle[twoside,fleqn,espcrc2,psfig]{article}

\def\msun{$M_{\odot}$}

\newcommand{\AmS}{{\protect\the\textfont2
  A\kern-.1667em\lower.5ex\hbox{M}\kern-.125emS}}

\title{Monitoring the Galactic Microquasars with $RXTE$}

\author{Ronald A. Remillard
and Edward H. Morgan\address{Center for Space Research,\\
        Massachusetts Institute of Technology,\\ 
        Cambridge MA 02139 USA}}

\begin{document}

\begin{abstract}
The Galactic microquasars GRS1915+105 and GRO J1655-40 were targets
for dedicated monitoring programs with $RXTE$ throughout 1996 and
1997. We review recent results with particular attention to the X-ray
QPOs seen in these sources. We confirm the reappearance of 67 Hz
oscillations in GRS1915+105 in several different spectral states. Six
intervals, each of 1 to 6 months duration, yield QPO detections with a
central frequency of $66.7 \pm 1.4$ Hz and a FWHM of $3.4 \pm 1.0$
Hz. Comparing these observing intervals, the average X-ray luminosity
varies by a factor of 4 with no significant change in the central QPO
frequency. GRS1915+105 has been active in X-rays since 1992, and it
shows no signs of evolving toward quiescence. In contrast,
GRO J1655-40, which began its most recent X-ray outburst on 1996 April
25, is seen to fade below 20 mCrab on 1997 Aug 17. There are no
reappearances of the 300 Hz QPO during the entire second half of the
X-ray outburst; during this time the X-ray spectrum is dominated by
the soft, thermal component. There are still several models seeking to
explain the high frequency QPOs in microquasars. While it is likely
that the cause invokes some effect of General Relativity, the correct
model and the details of the emission mechanism remain uncertain.
\end{abstract}

\maketitle

\section{INTRODUCTION}

Two Galactic X-ray sources known to produce relativistic radio jets
are GRS 1915+105 \cite{Mir94} and GRO J1655-40
\cite{Tin95,Hje95}. Optical observations of GRO J1655-40 have provided
dynamical evidence for a 7 \msun black hole \cite{Oro97} in a 2.6 day
binary orbit with a $\sim$F4 IV companion star. GRS1915+105 is
presumed to be a black hole binary, based on its X-ray high luminosity
and similarities with GRO J1655-40. However, a direct measurement of
the motion of its companion star has been prevented by interstellar
extinction, which limits optical/IR studies of GRS1915+105 to
wavelengths $> 1$ micron \cite{Mir94}. While each source was active
at radio frequencies, H I absorption measurements were combined with
Galactic rotation models to derive distance measurements of 12.5 kpc
and 3.2 kpc for GRS 1915+105 \cite{Mir94} and GRO J1655-40
\cite{Tin95}, respectively.

\begin{figure*}
\centerline{\psfig{figure=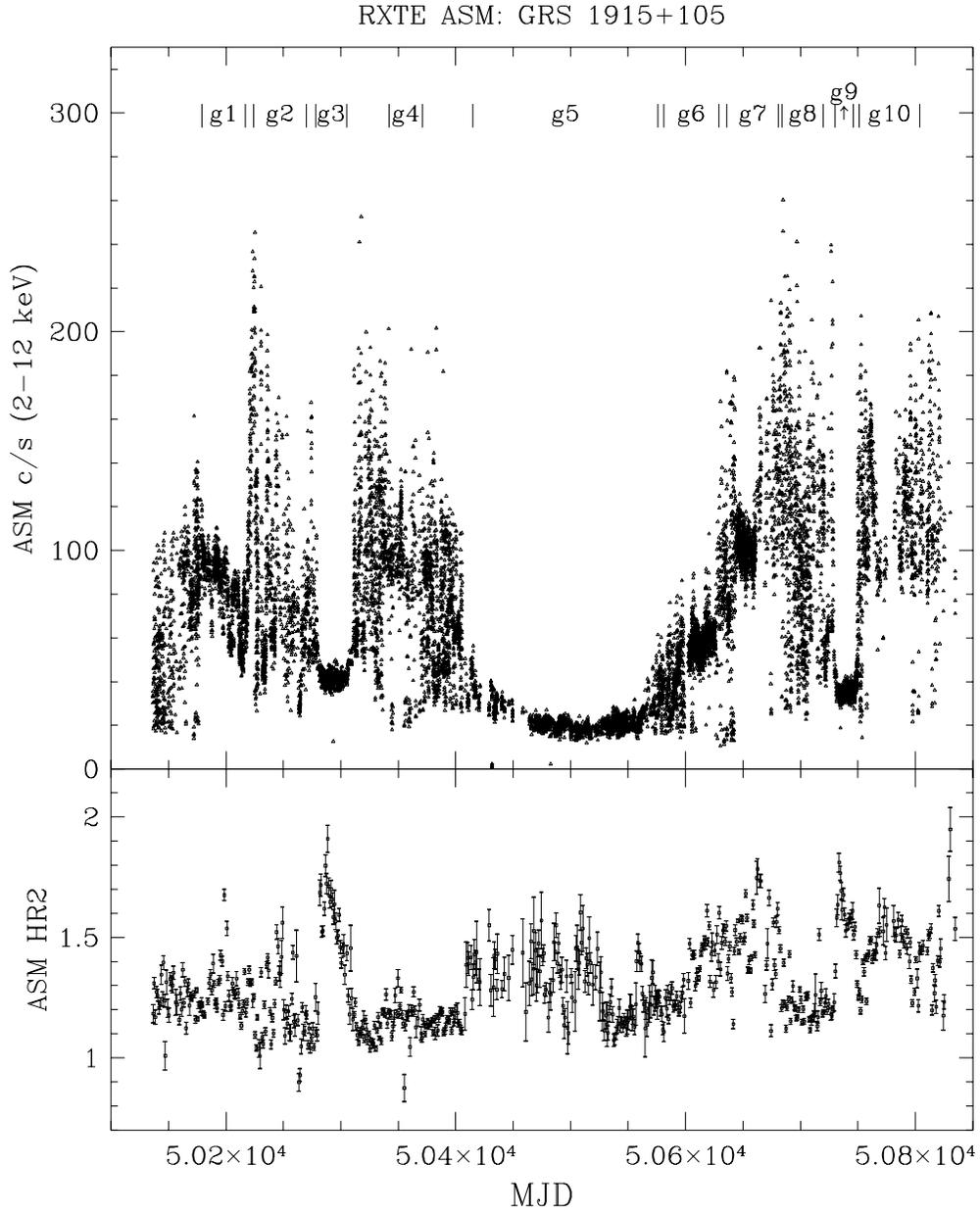,width=16cm,height=16cm} }
\caption{ASM light curve (2--12 keV) of GRS1915+105 for 1996 and
1997. The Crab Nebula, for reference, yields 75.5 ASM c/s. The ASM
hardness ratio, $HR2$ is defined as the count rate in the 5--12 keV
band relative to the rate in the 3--5 keV band. The time intervals
that correspond with our groups of combined X-ray power spectra (see
Table 1) are shown above the light curve.}
\label{fig:asm19}
\end{figure*}

GRS 1915+105 is a transient X-ray source, and the BATSE light curve
(20--100 keV) indicates that bright X-ray emission began during May
1992 \cite{Har97}. Before the launch of the $Rossi ~X$-$ray ~Timing
~Explorer$ ($RXTE$), observations in soft X-rays were sporadic, and
GRS1915+105 may have persisted as a bright source in soft X-rays since
1992. When the All Sky Monitor (ASM) on $RXTE$ established regular
coverage on 1996 Feb 22, the source was bright and highly variable,
and it has remained so throughout 1996 and 1997. The ASM light curve,
which is shown in Figure~\ref{fig:asm19}, illustrates both the extent
of the intensity variations and also the repetitive character of
particular variability patterns. The early ASM light curve was used to
initiate $RXTE$ pointed observations (PCA and HEXTE instruments),
which began on 1996 April 6.  Since then the source has been observed
once or twice per week, and most of the data are available in a public
archive. At the higher time resolution provided by PCA light curves,
there are again dramatic and repetitive patterns of variations
\cite{Gre96}.  These results are one of the extraordinary chapters in
the history of high-energy astronomy.

Fourier analyses of the first 31 PCA observations \cite{Mor97} of
GRS1915+105 revealed 3 different types of oscillations: a
quasi-periodic oscillation (QPO) with a constant frequency of 67 Hz;
dynamic, low-frequency (0.05 to 10 Hz) QPO with a large variety
of amplitudes and widths; and complex, high-amplitude dip cycles
($10^{-3}$ to $10^{-1}$ Hz) that are related to the extreme X-ray
variations noted above. The combined characteristics of the power
spectra, light curves, and energy spectra were interpreted as
representing four different emission states \cite{Mor97}, none of
which resemble the canonical states of black hole binaries
\cite{Van95}.

The other microquasar, GRO J1655-40, was first detected with BATSE on
1994 July 27, and the correlation between hard X-ray activity and the
ejections of relativistic radio jets \cite{Har95} was an important
step in establishing the relationship between accretion changes and
the formation of jets. During late 1995 and early 1996, GRO J1655-40
entered a quiescent accretion state, permitting optical spectroscopy
of the companion star, which led to our knowledge of the binary
constituents and mass of the black hole~\cite{Oro97}, as noted above.

\begin{figure*}
\centerline{\psfig{figure=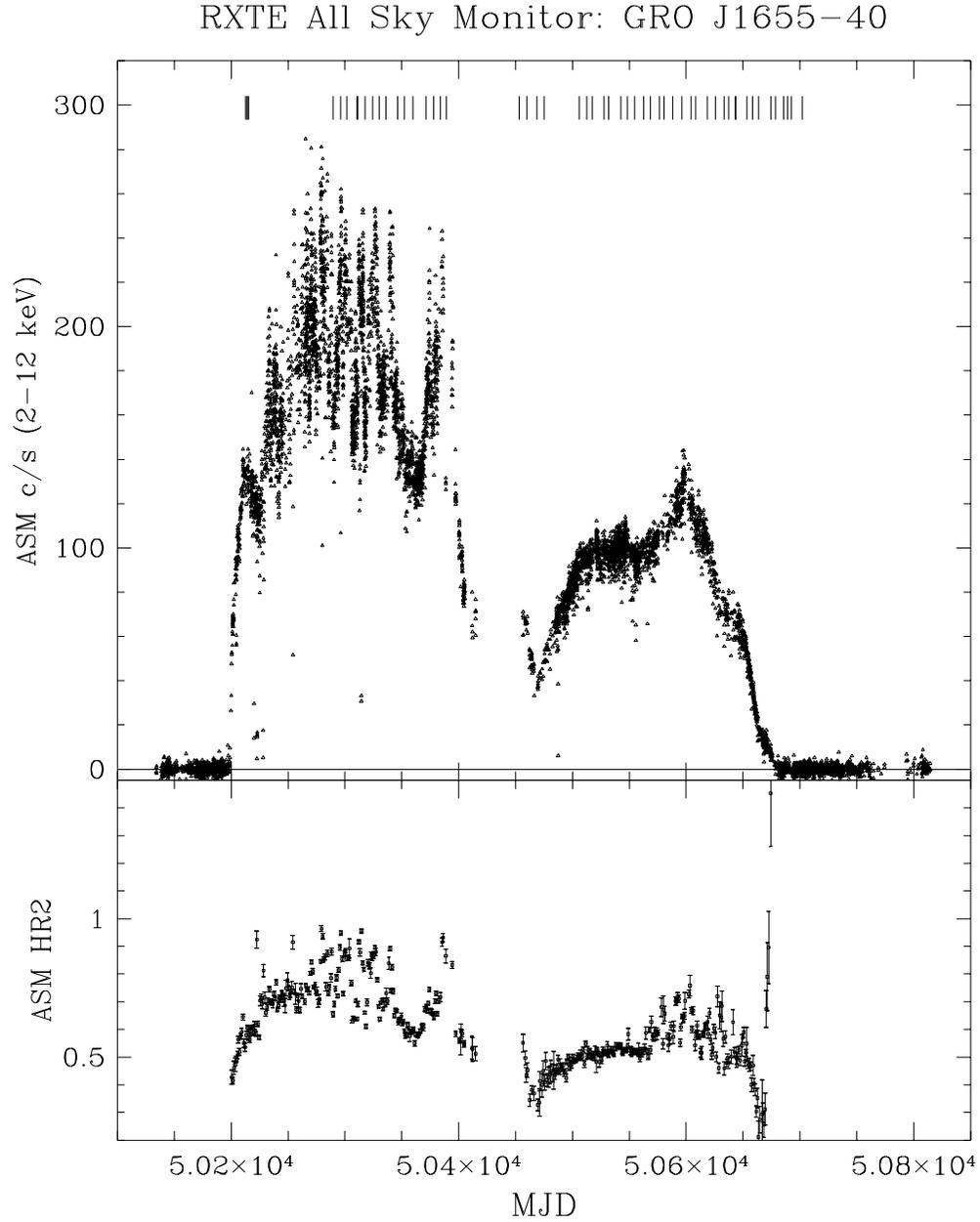,width=16cm,height=16cm} }
\caption{(top) ASM light curve (1.5--12 keV) of GRO J1655-40 for 1996
and 1997. The tick marks above the light curve show the times of RXTE
pointed observations, either from the public archive (1997) or our
guest observer program (1996). (bottom) The ASM hardness ratio, $HR2$
as defined previously.}
\label{fig:asm16}
\end{figure*}

The ASM recorded a renewed outburst from GRO J1655-40 \cite{Lev96}
that began on 1996 April 25. The ASM light curve is shown in Figure
~\ref{fig:asm16}. With great fortune a concurrent optical campaign
was in progress, and it was determined that optical brightening
preceded the X-ray turn-on by 6 days, beginning first in the I band
and then accelerating rapidly in the B and V bands. These results
provide concrete evidence favoring the accretion disk instability as
the cause of the X-ray nova episode.

The $RXTE$ observations of GRO J1655-40 indicate a more stable form of
accretion. X-ray spectral variations (see Fig.~\ref{fig:asm16})
resemble the canonical ``soft/high'' and ``very high'' states in black
hole binaries \cite{Rem98,Van95}. There are X-ray QPOs in the range of
8--22 Hz, and there is also a transient, high-frequency QPO at 300
Hz~\cite{Rem98}. This QPO is detected only when the X-ray power-law
component reaches its maximum strength.

The efforts to explain the 67 Hz QPO in GRS1915+105 and the 300 Hz QPO
in GRO J1655-40 commonly invoke effects rooted in General Relativity
(GR). There are at least 4 proposed mechanisms that relate the QPO
frequency to a natural time scale of the inner accretion disk in a
black hole binary.  These are: the last stable orbit
\cite{Sha83,Mor97}, diskoseismic oscillations \cite{Per97,Now97},
frame dragging \cite{Cui98}, and an oscillation in the centrifugal
barrier \cite{Tit98}. The physics of all of these phenomena invokes GR
effects in the inner accretion disk. It has also been proposed that
the high frequency QPOs may be caused by an inertial-acoustic
instability in the disk \cite{Che95} (with non-GR origin), although
the oscillation in GRO J1655-40 would extend this application to
higher frequencies than had been argued previously.

In this paper we advertise some recent work that associates jet
formation in GRS1915+105 with features in the X-ray light curve.  We
then turn to the topic of X-ray QPOs. New results are presented on the
reappearance of 67 Hz oscillations in GRS1915+105.  Finally we
describe the various QPO tracks that appear in GRO J1655-40, and we
explain how they behave in response to the strength of the power-law
component in the X-ray spectrum.

\section{CLUES FOR THE ORIGIN OF JETS IN GRS1915+105}

Several groups have combined X-ray, radio, and/or infrared
observations of GRS 1915+105 to probe the properties of jet formation
and relate the ejection events to features in the X-ray light curves.
Infrared jets were discovered \cite{Sam96}, and infrared flares were
seen to occur after radio flares\cite{Fen97,Mir97}. These
investigations provide solid evidence that the infrared flares
represent synchrotron emission from rapidly evolving jets.

It has been further demonstrated that the radio, infrared, and X-ray
bands occasionally show strong oscillations with a quasiperiodic time
scale of 20--40 min \cite{Rod97,Fen97,Eik98,Poo98}. In perhaps the
most impressive of these studies to date, there were a series of
infrared flares (with 20 min recurrence time), and in six of six
possible cases the flares were seen to follow dramatic dipping cycles
in the X-ray light curve. Since these dips have been analyzed as
representing the disappearance of the thermal X-ray emission from the
inner disk \cite{Bel97a,Bel97b}, the infrared/X-ray correlation shows
that the jet material originates in the inner accretion
disk\cite{Eik98}.  Another conclusion drawn from the recent
X-ray/radio/infrared studies is that there is a wide distribution of
``baby jets'' in which quantized impulses appear at $\sim30$ min
intervals.  The radio strength of these events is one to three orders
of magnitude below the levels of the superluminal outbursts of 1994
\cite{Poo98,Mir94}.

We expect that $RXTE$ will continue to support multifrequency
observations of GRS1915+105 during 1998. There are opportunities for
further analysis to characterize the distribution and expansion times
of the jets, analyze the infrared and radio spectra of these events,
and study the details of the X-ray light curve in the effort to
constrain the physics of the trigger mechanism.

\section{67 HZ OSCILLATIONS IN GRS1915+105}

There have been many observations of GRS1915+105 with $RXTE$ since the
six (1996 April 6--June 11) that provided detections of QPO at 67 Hz
\cite{Mor97}. Given the importance of this QPO and also the variety of
emission states recorded for GRS1915+105 (see Figure~\ref{fig:asm19}),
we investigated the data archive for new detections of this QPO. We
adopted a global perspective, and we divided the $RXTE$ observations
into a sequence of X-ray state intervals, which we label as groups
``g1'' through ``g10'' in Figure~\ref{fig:asm19}. The groups were
selected with consideration of both the ASM light curve and the
characteristics of the PCA power spectra, and some observations
between the group boundaries were ignored as representing transition
states.

In Table~\ref{tab:67hz} we list the time intervals (cols. 2, 3) the
number of observations (col. 4), the X-ray state (col. 5), and the
average X-ray flux (in Crab units) for each group. The typical
observation has an exposure time of 10 ks. The X-ray state
description follows the convention of Morgan et al. \cite{Mor97},
which describes GRS1915+105 as being relatively steady and bright (B),
flaring (FL), chaotic (CH), or low-hard (LH).

\begin{table*}


\newlength{\digitwidth} \settowidth{\digitwidth}{\rm 0}
\catcode`?=\active \def?{\kern\digitwidth}
\caption{The 67 Hz QPO in GRS1915+105}
\label{tab:67hz}
\begin{tabular*}{\textwidth}{@{}l@{\extracolsep{\fill}}llrccccc}
\hline
group & start & end & obs & state & flux & freq. & FWHM & ampl. \\
\hline

1  & 1996 Apr 06 & 1996 May 14 &  7 & B     & 1.06 & 64.5 & 4.0 & 0.0069 \\
2  & 1996 May 21 & 1996 Jul 06 & 14 & FL    & 1.00 & 65.7 & 2.3 & 0.0022 \\
3  & 1996 Jul 14 & 1996 Aug 10 &  6 & LH    & 0.58 &      &      & \\
4  & 1996 Sep 16 & 1996 Oct 15 &  8 & B     & 1.01 & 67.6 & 1.5 & 0.0016 \\
5  & 1996 Nov 28 & 1997 May 08 & 28 & LH    & 0.31 & 68.3 & 2.3 & 0.0023 \\
6  & 1997 May 13 & 1997 Jun 30 & 18 & CH/B  & 0.64 &      &     & \\
7  & 1997 Jul 07 & 1997 Aug 21 & 17 & B     & 1.33 & 66.9 & 4.3 & 0.0039 \\
8  & 1997 Aug 24 & 1997 Sep 29 & 15 & CH/FL & 1.17 &      &     & \\
9  & 1997 Oct 09 & 1997 Oct 25 &  4 & LH    & 0.47 &      &     & \\
10 & 1997 Oct 30 & 1997 Dec 22 & 15 & FL    & 1.41 & 67.4 & 4.2 & 0.0035 \\
\hline
\end{tabular*}
\end{table*}

\begin{figure*}
\centerline{\psfig{figure=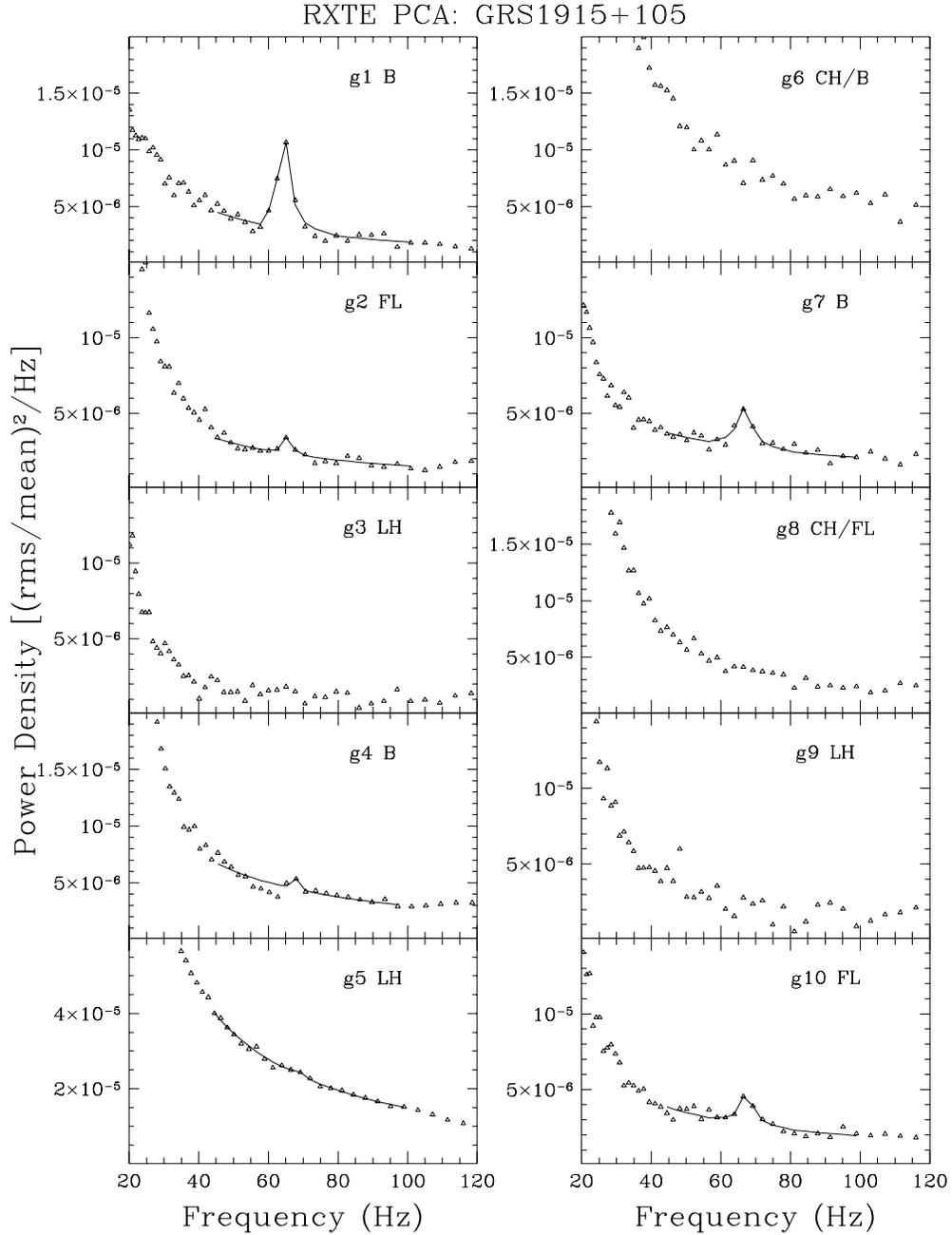,width=16cm,height=16cm} }
\caption{Average power density spectra in the range of 20--120 Hz for RXTE PCA observations of 1996 and 1997, combined in 10 groups. For the 6 cases in which a QPO is detected (see Table 1), the QPO fit is shown with a solid line.}
\label{fig:fit67hz}
\end{figure*}

We then combined the power spectra in each group, using the full
energy coverage of the PCA instrument.  We fit the results for a power
continuum (with a power-law function) and a QPO feature (with a
Lorentian profile) over the range of 40--120 Hz. We emphasize that the
location of the central QPO frequency is free to wander within this
frequency interval. The average power spectra for the 10 groups
(linear units) and the QPO fits for 6 cases are shown in
Figure~\ref{fig:fit67hz}.

The results derived from this analysis are listed in
Table~\ref{tab:67hz}.  The central QPO frequency is given in col. 7,
and there is a narrow distribution of $66.7 \pm 1.4$ Hz. The QPO FWHM
values (col. 8) have a mean value of $3.4 \pm 1.0$ Hz. Comparing these
observing intervals, we conclude that the average X-ray luminosity of
GRS1915+105 may vary by a factor of 4 with no significant change in
the characteristics of the 67 Hz QPO.

The integrated QPO amplitude is given in col. 9.  The amplitudes (like
the power spectra in Figure~\ref{fig:fit67hz}) are normalized by the
mean X-ray count rate for GRS1915+105. The integrated power in the 67
Hz QPO is in the range of 0.2\%--0.7\% of the mean X-ray flux.

The results for group 5 are particularly noteworthy. During this
period the source was in the low-hard state for a long time (see
Figure~\ref{fig:asm19}. The PCA light curves in 1 s time bins show
variations limited to moderate flickering, with rms variations $\sim
10$\%. However the continuum power at 40--120 Hz is relatively high
during this interval (see Figure~\ref{fig:fit67hz}). The large number
of observations in group 5 partially compensates for the losses in
statistical sesitivity to QPO detection due to lower count rate and
elevated continuum power. Nevertheless the QPO search does find a
small feature that is consistent in frequency (68.3 Hz), width (2.3
Hz), and amplitude (0.23\%) with the other detections. We estimate that
the uncertainty in the amplitude is 0.09\%, so that the detection of
the 67 Hz QPO in group 5 has a signigicance of 2.6 $\sigma$.  For the
4 groups that do not yield QPOs in the range of 40--120 Hz, the
uncertainties are slightly larger, and we cannot exclude the
possibility that GRS1915+105 is $always$ emitting X-ray QPOs at 67 Hz
with amplitudes in the range of 0.1\% or larger.

There are yet many avenues for further investigation of this QPO,
e.g. time lags at 67 Hz, analysis of the energy spectrum for the
groups with positive QPO detection, and segregation of data with
alternative schemes such as the phases of jet-related dipping cycles.
All of these topics will be pursued during the next several months.

\section{QPOs in GRO J1655-40}

We have conducted similar analyses of PCA power spectra for individual
observations of GRO J1655-40. As reported previously \cite{Rem98},
there are transient QPOs in the range of 8--30 Hz and there is a high
frequency QPO near 300 Hz. All of these QPOs are associated with the
strength of the power-law component. With respect to
Figure~\ref{fig:asm16}, the QPOs at 8--30 Hz appear when observations have
hard spectra that correspond with ASM HR2 values above 0.8, while
the 300 Hz QPO is significant only when the combine the power spectra
for the 7 ``hardest'' observations made with the PCA (1996 August and
October). 

\begin{figure*}
\centerline{\psfig{figure=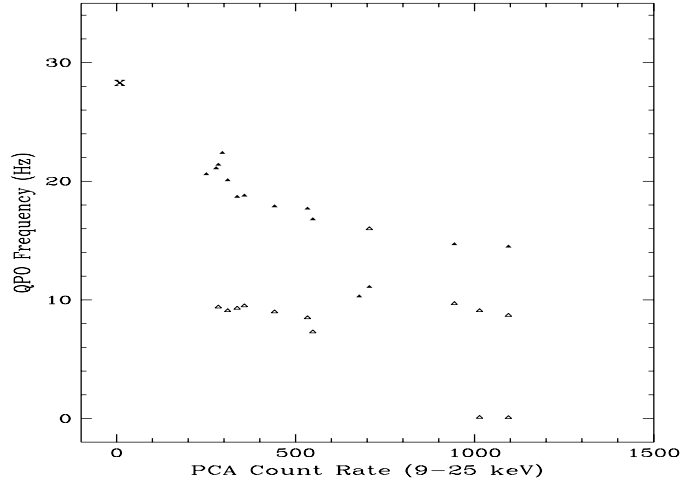,width=10cm,height=5.5cm} }
\caption{The central frequency of X-ray QPOs in GRO J1655-40 as a function of the PCA count rate above 13 keV. The open triangles represent broad QPOs, while the solid triangles represent narrow ones.}
\label{fig:qpo16}
\end{figure*}

We fit the individual PCA power spectra for power continuum and QPOs,
as described above, using frequency windows of 0.02--2 Hz and 5--50
Hz. In Figure~\ref{fig:qpo16} we show the central QPO frequencies as a
function of the source count rate in the PCA energy channels above 13
keV (or above channel 35).  We use on open triangle for narrow
QPOs ($\nu / \delta\nu > 5$) and the ``*'' symbol for broad QPOs
($\nu / \delta\nu < 4$). In some observations, both narrow and broad
QPOs appear in the same power spectrum (i.e. one 10 ks observation).
The ``x'' symbol shows a narrow and weak QPO derived from the average
power spectrum obtained during the 1997 PCA observations (MJD interval
50500--50650).

The results in Figure~\ref{fig:qpo16} show that the low-frequency QPOs
in GRO J1655-40 are organized in three tracks. A broad QPO appears to
be stationary near 8 Hz, while the narrow QPOs shifts to lower
frequency as the hard X-ray flux increases. The QPO derived from the
sum of 1997 observations appears to be a simple extension of this
narrow QPO track, occurring when the X-ray flux above 13 keV is nearly
zero.  Very low frequency QPO (0.085 and 0.11 Hz) are seen on two
occasions when the hard X-ray flux is near maximum.  These QPO coexist
with 300 Hz QPO, and they are reminiscent of the 0.067 Hz QPOs in
GRS1915+105. We speculate that the 0.1 Hz QPOs appear near the
threshold of the chaotic light curves manifest in GRS1915+105.
GRO1655-40 approaches this threshhold but does not cross the line into
unstable light curves during the 1996-1997 outburst.

In Figure~\ref{fig:asm16} we see that GRO J1655-40 fades below 20
mCrab on 1997 Aug 17. Whether there will be a renewed outburst in
1998 is anyone's guess, but the ASM will surely be monitoring this
source for any signs of X-ray activity.


\begin{thebibliography}{9}
\bibitem{Bel97a} T. Belloni, M. Mendez, A.R. King, M. van der Klis, \& J. van Paradijs, Ap. J. Letters 479 (1997) L145.
\bibitem{Bel97b} T. Belloni, M. Mendez, A.R. King, M. van der Klis, \& J. van Paradijs, Ap. J. Letters 488 (1997) L109.
\bibitem{Che97} W. Chen, C.R. Shrader, \& M. Livio, Ap. J. 491 (1997) 312.
\bibitem{Che95} X. Chen \& R.E. Taam, Ap. J. 441 (1995) 354.
\bibitem{Eik98} S.S. Eikenberry, K. Matthews, E.H. Morgan, R.A. Remillard, \& R.W. Nelson, Ap. J. Letters, (1998) in press.
\bibitem{Cui98} W. Cui, S.N. Zhang, \& W. Chen, Ap. J. Letters, 492 (1998) L53.
\bibitem{Gre96} J. Greiner, E. Morgan, \& R. Remillard, Ap. J. Letters 473 (1996) L107.
\bibitem{Fen97} R.P. Fender, G.G. Pooley, C. Brocksopp, \& S.J. Newell, MNRAS, 290 (1997) L65.
\bibitem{Har95} B.A. Harmon, et al., Nature 374 (1995) 703.
\bibitem{Har97} B.A. Harmon, K.J. Deak, W.S. Paciesas, S.N. Zhang, C.R. Robinson, E. Gerard, L.F. Rodriguez, \& I.F. Mirabel, Ap. J. Letters, 477 (1997) L85.
\bibitem{Hje95} R.M. Hjellming \& M. Rupen, Nature, 375 (1995) 464.
\bibitem{Lev96} A.M. Levine, H. Bradt, W. Cui, J. Jernigan, E.H. Morgan, R.A. Remillard, R.E. Shirey, \& D.A. Smith, Ap. J. Letters, 469 (1996) L33.
\bibitem{Mir94} I.F. Mirabel \& L.F. Rodriguez,  Nature 371 (1994) 46.
\bibitem{Mir97} I.F. Mirabel, L.F. Rodriguez, S. Chaty, M. Sauvage, E. Gerard, P.A. Duc, A. Castro-Tirado, \& P. Callanan, Ap. J. Letters 472 (1997) L111.
\bibitem{Mor97} E.H. Morgan, R.A. Remillard, \& J. Greiner, Ap. J. 482 (1997) 993.
\bibitem{Now97}  M.A. Nowak,  R.V. Wagoner, M.C. Begelman, \& D.E. Lehr, Ap. J. Letters 477 (1997) L91.
\bibitem{Oro97} J.A. Orosz \& C.D. Bailyn, Ap, J. 477 (1997) 876.
\bibitem{Orr97} J.A. Orosz, R.A. Remillard, C.D. Bailyn, \& J.E. McClintock,  E. Ap. J. Letters 478 (1997) L83.
\bibitem{Per97} C.A. Perez, A.S. Silbergleit, R.V. Wagoner, \& D.E. Lehr, Ap, J. 476 (1997) 589.
\bibitem{Poo98} G.G. Pooley \& R.P. Fender, MNRAS, (1998) in press.
\bibitem{Rem98} R.A. Remillard, E.H. Morgan, J.E. McClintock, C.D. Bailyn, \& J.A. Orosz, Procs. Texas Symposium, eds. Olinto, Frieman, \& Schramm, World Scientific Press, (1998) in press; astro-ph/9705064.
\bibitem{Rod97} L.F. Rodriguez \& I.F. Mirabel, Ap. J. Letters, 474 (1997) L123.
\bibitem{Sam96} B.J. Sams, A. Eckart, \& R. Sunyaev, Nature 382 (1996) 47.
\bibitem{Sha83} S.L. Shapiro \& S.A. Teukolsky, $Black ~Holes, White ~Dwarfs, ~and ~Neutron ~Stars$, Wiley \& Sons (1983).
\bibitem{Tin95} S.J. Tingay et al., Nature 374 (1995) 141.
\bibitem{Tit98} L. Titarchuk, I. Lapidus, \& A. Muslimov, Ap. J.. submitted.
\bibitem{Van95} van der Klis, M. 1995, in $X-ray ~Binaries$, eds. W. Lewin, J. van Paradijs, \& E. van den Heuvel, Cambridge University Press, p. 252.
\end{thebibliography}
\end{document}